\documentclass[rnote]{aa}
\usepackage[varg]{txfonts}
\usepackage{graphicx}
\usepackage{amstext}

\begin{document}

\title{San Pedro M\'artir observations of microvariability in obscured quasars }
\titlerunning{Microvariability in type 2 QSOs}

\author{J. Polednikova\inst{1,2} \and A. Ederoclite\inst{3} \and J. Cepa\inst{1,2} \and J.A. de Diego\inst{4,5} \and  J.I. Gonz\'alez-Serrano\inst{6} \and A. Bongiovanni\inst{1,2} \and I. Oteo\inst{7,8} \and A.M. P\'erez Garc\'ia\inst{1,2,9} \and R. P\'erez-Mart\'inez\inst{10, 11} \and I. Pintos-Castro\inst{1,2,12} \and M. Ram\'on-P\'erez\inst{1,2} \and M. S\'anchez-Portal\inst{11,13}}

\authorrunning{Polednikova et al.}


\institute{Instituto de Astrof\'isica de Canarias, C/Via Lactea s/n, La Laguna, 38205 Spain 
\and Departamento de Astrof\'isica, Universidad de La Laguna, La Laguna, Spain
\and Centro de Estudios de F\'isica del Cosmos de Arag\'on, Teruel, Spain
\and Instituto de Astronom\'ia, Universidad Aut\'onoma de M\'exico, 04310, M\'exico D.F., M\'exico
\and Instituto de Astrof\'isia de Canarias - Universidad de La Laguna, CEI Canarias: Campus Atl\'antico Tricontinental, La Laguna, 38205 Spain 
\and Instituto de F\'isica de Cantabria (CSIC-Universidad de Cantabria), Santander, Spain
\and Institute for Astronomy, University of Edinburgh, Royal Observatory, Blackford Hill, Edinburgh EH9 3HJ
\and European Southern Observatory, Karl-Schwarzschild-Str. 2, 85748 Garching, Germany
\and ASPID Association, Ap. correos 412, La Laguna, Spain
\and XMM/Newton Science Operations Centre, ESAC/ESA. Villanueva de la Cañada, Madrid, Spain
\and Ingeniería de Sistemas para la Defensa de España (Isdefe), Madrid, Spain
\and Centro de Astrobiolog\'ia, INTA-CSIC, Villanueva de la Ca\~nada, Madrid, Spain
\and Herschel Science Centre, ESAC/ESA, Villanueva de la Cañada, Madrid, Spain
}





\abstract{

Fast brightness variations are a unique tool to probe the innermost regions of  active galactic nuclei (AGN). These variations are called microvariability or intra-night variability, and this phenomenon has been monitored in samples of blazars and unobscured AGNs. Detecting optical microvariations in targets hidden by the obscuring torus is a challenging task because the region responsible for the variations is hidden from our sight. 
However, there have been reports of fast variations in obscured Seyfert galaxies in X-rays, which rises the question whether microvariations can also be detected in obscured AGNs in the optical regime. Because the expected variations are very small and can easily be lost within the noise, the analysis requires a statistical approach. We report the use of a one-way analysis of variance, ANOVA, with which we searched for microvariability. ANOVA was successfully employed in  previous studies of unobscured AGNs. 
As a result, we found microvariable events during three observing blocks: in two we observed the same object (Mrk 477), and in
another, J0759+5050. The results on Mrk\,477 confirm previous findings.  However, since Mrk\,477 is quite a peculiar target with hidden broad-line regions, we cannot rule out the possibility that we have serendipitously chosen a target prone to variations.}

\maketitle

\section{Introduction}

The different spectral features of AGNs can be explained by the different orientation of the object along the observer's line of sight. Type 2 AGNs are oriented edge-on, and therefore the view of the central engine is blocked by the obscuring torus. Type 2 AGNs can furthermore be divided according to their luminosity. While Seyfert 2 galaxies are low-luminosity type 2 AGNs, type 2 quasars are the high-luminosity counterparts. Determining the luminosity of obscured AGNs is no trivial task because the dust obscuring torus effectively blocks our view. Therefore \citet{rey08} proposed a proxy, where obscured AGNs with $L_{\rm{[OIII]}} \geq 10^{8.3}L_{\odot}$ are classified as type 2 quasi-stellar objects
(QSOs).

Microvariability (or intra-night variability) is a phenomenon of variations with a low amplitude (hundredths of a magnitude) over a short timescale (minutes to hours). The first reports of microvariations were made shortly after the discovery of quasars \citep{mas}, but the reported microvariations were considered unreliable because of technical difficulties. The first reliable detections of microvariations were reported by \citet{mil89} in blazars, which are extremely powerful radio sources.

Microvariability must originate in a region of some light minutes in size to agree with causality arguments. We can assume that this region is located near the center associated with the most violent physical phenomena. Microvariability may arise either from instabilities in the jet or in the accretion disk, or both  \citep{ram, ram10}, and therefore provides a unique opportunity to study the unresolved inner regions. These regions are hidden in type 2 QSOs, and the contribution of the host galaxy to the total flux cannot be neglected, thus these objects are unlikely to show microvariations.

This paper is organized as follows: in Sect. 2, we describe the sample, observations, and data treatment. Section 3 is dedicated to the analysis of the data and the brief description of the statistical test used. In Sect. 4 we summarize the results and discuss them along with possible explanations in Sect. 5. At the end, we provide short conclusions.

\section{Sample selection, observations, and data reduction}

\subsection{Sample}

Our aim was to obtain observations of a homogeneous set of obscured type 2 quasars. We have chosen type 2 quasars over Seyfert 2 galaxies because the contamination of the host galaxy is lower than in the Seyfert 2 galaxies as a result of the higher intrinsic brightness of the nuclei in type 2 QSOs. 
The sample was selected on the basis of brightness. It was chosen from the catalog provided by \citet{rey08}, which is based on
the Sloan Digital Sky Survey Data Release 6 \citep{dr6}. This catalog consists of 887 type 2 quasars selected based on [OIII] emission line luminosity. It is the most complete collection of  type 2 QSOs to date. The limiting redshift of the catalog is $z = 0.83$. The catalog covers 80\% of the SDSS DR6 spectroscopic database. Our targets were chosen to be bright ($g<17$ mag) to achieve accurate photometric measurements with the telescope.
There are four objects fulfilling our requirements in the catalog. All of them are observable
from the northern hemisphere. We were able to observe three of them. The fourth target met with telescope-pointing problems. The list of the observed targets and their properties can be found in Table \ref{tbl-1}. 
All the chosen targets are at $z<0.1$. Thanks to this, we can compare similar properties between the sample objects because the same phenomenon falls into similar broadband filter.

\begin{table}
\begin{center}
\caption{Properties of the observed targets from the catalog
of \citet{rey08}. Right ascension and declination are in J2000. Brightness in V filter was transformed from SDSS filters. The values correspond to the catalog of quasars and active nuclei: 13th edition \citep{ver}.\label{tbl-1}}
\begin{tabular}{ccccc}
\hline
Target & R.A. & dec & $V_{ mag}$ & $z$ \\
\hline
J0759+5050 & 07:59:40 & +50:50:24 & 16.59 & 0.06 \\
J1430+1339 & 14:30:29 & +13:39:12 & 17.64 & 0.09 \\
Mrk 477 & 14:40:38& +53:30:19 & 15.03 & 0.04 \\
\hline
\end{tabular}
\end{center}
\end{table}

\subsection{Observations and data reduction}

Our sample was observed between March 28 and March 31, 2011, with the SITe 4 detector mounted on the  1.5 m telescope at San Pedro M\'artir
observatory in Baja California, M\'exico.  We employed the same observing strategy as
in previous works \citep{die98,ram}. All targets were observed in the Johnson V filter.
Exposure time for each target was set to 60 seconds, providing a singal-to-noise ratio (S/N) $> 100$. 
This corresponds to a precision of 0.01 mag. The detector was binned (2x2) to allow fast readout with low noise. We observed all the targets with less than 1.3 airmass, and all of our observing nights were dark. We checked the colors of the stars that we have used for the analysis. \citet{car} and \citet{sta} proposed that the airmass effect is negligible as long as the color difference is $g-r < 1.5$. We provide the color information for our targets and stars in the Table \ref{tbl-2}.

\begin{table}
\begin{center}
\caption{Colors of the targets, reference stars, and comparison stars for every observed field. The values are adopted from the SDSS data release 12 \label{tbl-2}.}
\begin{tabular}{c|ccc}
\hline
Field & star ID & $g$ & $g-r$ \\
\hline
J0759+5050 & & 16.09 & 0.31\\
                & star1&15.58& 0.44\\
                &star2&13.65&1.18\\
J1430+1339&&15.80&0.37\\
                &star1&16.68&1.1\\
                &star2&15.01&1.21\\
Mrk477&&14.84&0.14\\
                &star1&15.01&0.60\\
                &star2&13.23&-1.05\\
\hline
\end{tabular}
\end{center}
\end{table}

Bias subtraction and flat-fielding were performed using the standard tasks from the IRAF data 
reduction software\footnote{Image Reduction and Analysis Facility (http://iraf.noao.edu/).}.
Cosmic rays were removed using L.A.Cosmic\footnote{Laplacian cosmic ray identification/ (http://www.astro.yale.edu/dokkum/lacosmic/).} by \citet{vad}.
Aperture photometry was performed using SExtractor 2.8.6\footnote{Source Extractor (http://www.astromatic.net/software/sextractor).} \citep{sex}. We used fixed apertures for 
both the target and the stars in the field. We made sure to enclose all stars and host galaxies detectable in our images. This is important because the contribution can be variable for bright galaxies as a result of intra-night fluctuations in the atmospheric seeing, as pointed out by \citet{car}. According to \citet{cel}, an aperture that would include the whole host galaxy can account for the seeing variations. For stars, we used apertures of 25 px  (corresponding to 6.32 arcsec) diameter, while for the AGNs we used a 40 px (10.12 arcsec) diameter aperture. This is roughly four times the full width at half maximum (FWHM) of the objects.  Hence we assume that with the apertures that we have chosen, the contribution of the host galaxy is constant over time and therefore is not causing any additional variations. 

\section{Analysis}
We took advantage of the presence of other stars in the field
to perform differential photometry. We used one star as a reference and another one for comparison to ensure the stability conditions. The left columns of Fig. 1 show the stability of the field stars during the observations. 

According to \citet{die10}, there are two statistical tests suitable to search for microvariability, namely an F-test and a one-way analysis of variance; ANOVA. ANOVA surpasses F-test in power in general, but an enhanced F-test \citep{die14} can provide more statistical power than ANOVA. Unfortunately, the enhanced F-test takes advantage of the presence of many objects in the same field as the target to provide calibrating stars. Figure 13 of \citet{die14} shows that ANOVA is the most suitable test for our needs. 

\begin{figure}
\centering
\includegraphics[width=9cm]{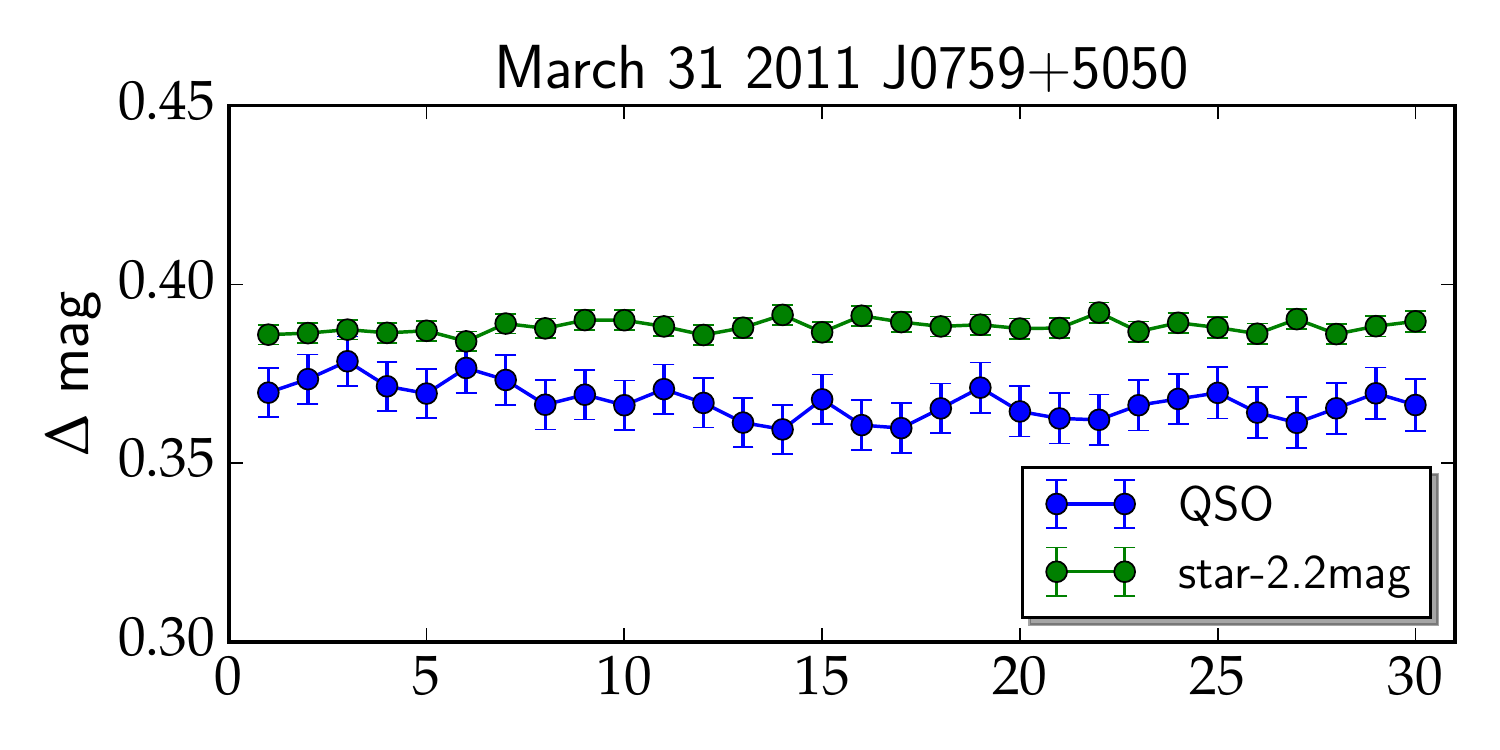}
\caption{Differential light curve with errors computed as standard deviations within the sets of five observations for one of the targets and a comparison star. A constant is subtracted from the star to fit it into one plot\label{Fig2}.}
\end{figure}

The observing strategy was specifically chosen with the aim to use ANOVA. For each object, we divided the data from every night into groups of five data points, following the strategy of \citet{die98}. We computed the variance in each of the groups and the variance among the groups. 
We largely benefited from the fact that ANOVA does not rely on the error estimation from the photometric software. The errors are computed from the standard deviation within each group of five observations. Light curves with error bars estimated from ANOVA for every first observing night of each target are provided in Fig. \ref{Fig2}.

\begin{table*}
\begin{center}
\caption{Results of microvariability in type 2 quasars. The second column describes the difference we used to create the differential light curve, where $m_{QSO}$ stands for the target, $m_{ref}$ stands for the reference star, and $m_{comp}$ stands for the comparison star used to control the stability of the stellar light curve. $N$ stands for the number of groups used as an input for ANOVA. In the fifth column, we provide the probability of obtaining the result if the null hypothesis is confirmed. The level of significance is $\alpha = 0.001$).\label{tbl-3}} 
\begin{tabular}{cccccc}
\hline
Target & DLC & Date & N & Probability & Microvariability ($\alpha = 0.001$)\\
\hline
J0759+5050 & $m_{QSO} - m_{ref}$ & Mar 28 2011 & 30 & 0.0975 & N\\
 & $m_{comp} - m_{ref}$ & Mar 28 2011 & 30 & 0.132 & N \\
 & $m_{QSO} - m_{ref}$ & Mar 30 2011 & 20 & 0.073 & N\\
 & $m_{comp} - m_{ref}$ & Mar 30 2011 & 20 & 0.146 & N\\
& $m_{QSO} - m_{ref}$ & Mar 31 2011 & 30 & 5.5E-04 & Y\\
 & $m_{comp} - m_{ref}$ & Mar 31 2011 & 30 & 0.327 & N\\
J1430+1339\tablefootmark{a} & $m_{QSO} - m_{ref}$ & Mar 28 2011 & 25 & 0.0105 & \dots\\
 & $m_{comp} - m_{ref}$ & Mar 28 2011 & 25 & 0.027 &\dots\\
 & $m_{QSO} - m_{ref}$ & Mar 30 2011 & 12 & 0.391 & N\\
 & $m_{comp} - m_{ref}$ & Mar 30 2011 & 12 & 0.436 & N\\
Mrk 477 & $m_{QSO} - m_{ref}$ & Mar 29 2011 & 30 & 1E-04 & Y\\
 & $m_{comp} - m_{ref}$ & Mar 29 2011 & 30 & 0.114 & N\\
 &  $m_{QSO} - m_{ref}$ & Mar 31 2011 & 20 & 4.95E-05 & Y\\
 & $m_{comp} - m_{ref}$ & Mar 31 2011 & 20 & 0.102 & N\\
\hline
\end{tabular}

\tablefoottext{a}{Target J1430+1339 has the same probabilities for stars and target, all of them low. The lack of stars in the field does not allow conclusions based on this observation.}
\end{center}
\end{table*}

\section{Results}

\begin{figure*}
\centering
\includegraphics[width=17cm]{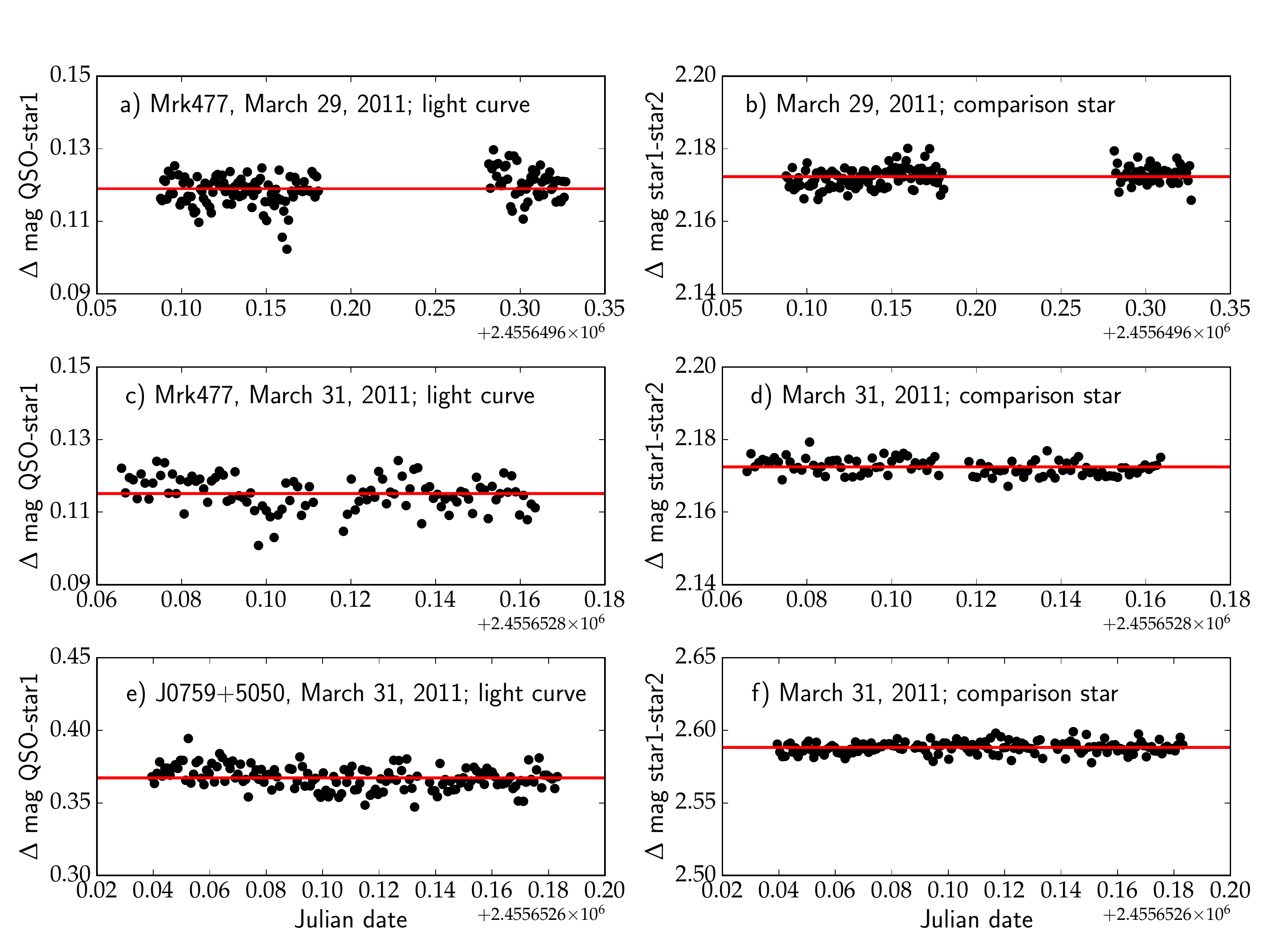}
\caption{Differential light curves for Mrk 477 (panels $a - d$) and J0759+5050 (panels $e - f$) for the nights with detected variability. Panels $a, c,$ and $e$ represent $m_{QSO} - m_{ref}$, while panels $b,d,$ and $f$ represent $m_{comp} - m_{ref}$.
The reference star light curves do not show any features, as expected. \label{Fig3}}
\end{figure*}

We have detected microvariability in two out of three targets that we observed in the run. It has been shown that there is a strong evidence, at the $\alpha = 0.001$ level of significance, of variability in the light curves of threee observations of type 2 quasars.  Below, we discuss the results for individual targets. We present light curves for the microvariable events in Fig. \ref{Fig3}. For the microvariable events, we also provide the microvariability amplitudes as defined by
 \citet{hew}, 

\begin{equation}
Amp = \sqrt{(A_{max} - A_{min})^2 - 2\sigma^2}.
\end{equation}

\textbf{J0759+5050} was observed during three nights.  The observations from March 30, 2011, was disturbed by clouds, so it is not strictly continuous. 

Microvariability was only detected during the observations carried out on March 31. The event was observed during the period between 3:21 and 5:42 UT.  The upper level of the amplitude is 0.04 mag, which is well above our estimated precision set by the high S/N. The variability amplitude is 4.6\%.  
Because the microvariations were only observed during one of the nights of the campaign, we conclude that we cannot confirm microvariability in this source. 

\textbf{J1430+1339} was observed on March 28 and March 30, 2011. During the two observing nights, both the target and the comparison star exhibited very contradictory results from ANOVA. We were unable to find a comparison star that was not variable at $\alpha = 0.001$, although the $p-\text{values}$ were likewise too low to confirm variability. Unfortunately, the field around this particular target limited the choice of comparison stars we could have used for ANOVA. In conclusion, there is not enough evidence of microvariability in this source. 

\textbf{Mrk 477} was observed on March 29 and March 31, 2011. The gap between the observations on March 29 was caused by a technical maintenance pause. Mrk 477 is the only target that showed microvariable behavior in both observing nights. The upper limit for variations is 0.027 between 5:55 UT and 7:02 UT and 0.023 mag between 8:46 UT and 10:02 UT (for the first and second observing night, respectively). Computed amplitudes of microvariability are 2.7\%and 2.3\% for the fisrt and second observing night, respectively. 
Based on our results, we conclude that Mrk 477 exhibits microvariations.

\section{Discussion}

We have detected microvariable events with a confidence level of $\alpha = 0.001$ during three observing blocks. The positive detections were made for J0759+5050 (one observing block out of three) and Mrk 477 (microvariability detected in two out of two observing blocks). J0759+5050 was only variable during one of the observing blocks, we were unable to confirm the variability behavior in the other observing nights. The amplitudes of the variations as computed using the approach by \citet{hew} are comparable to the other quantitative measurements of microvariability \citep{rom}.
It is highly unlikely that our measurements
are affected by systematics because possible effects would also affect the comparison stars, which we have detected as not variable. Because we observed during  dark nights, we can rule out any variation caused by significant change in the background level. We also controlled the airmass of the observations. Since the airmass does not rise above 1.3 airmasses, it is highly unlikely that any of the variations would be based on the color change of the source. Given that the target detected as variable is in the vicinity of other sources in the image, we constructed check graphs with different aperture sizes to validate that there is no spurious light from other sources propagating into our measurements. The apertures are sufficiently large to be unaffected by a change of the seeing.

Mrk 477, the brightest and closest of our sources, is the only one that was widely observed before our study. It was already reported to show optical microvariability by \citet{jan}, although the report lacks details and comments on robustness of the result and the statistics used. \citet{tra} reported detection of  hidden broad-line regions (HBLR) as a part of spectropolarimetric observations. \citet{tra} found that the degree of polarization is $P=1.2\%$ and the rotation of the position angle is $\Theta = 85^\circ$. The presence of HBLR might provide a clue about the mechanism behind the variability of the source. Until now, almost all the sources that show optical microvariability are type 1 AGNs, meaning
that they have a visible broad-line region, and we can assume that part of the variations originates there. Our detection of  variability would therefore be consistent with the detection of the HBLR reported by \citet{tra}. The mechanism that allows us to detect the variability remains unclear, although there are possible explanations.  

It is possible that the observed light originates close to the central engine and is then reflected on the corona of charged particles above the dusty torus. Polarimetric observations would be needed to confirm such a scenario. We expect that microvariations in obscured sources show small amplitudes, therefore this scenario would agree with our results. Nevertheless, we cannot confirm this statement as we would need to observe the source for an extended amount of time in polarized light. 


Mrk 477 is also the only target out of our sample that was detected in X-rays. It was reported as a Compton-thick source \citep{gil}. Recently, there have been reports of spectral variability in X-ray detected obscured sources \citep[most notably the Seyfert 1.8 galaxy NGC 1365;][]{ris09, bre12} on short timescales. The variations are caused by alternate covering and uncovering of the X-ray source by clouds in the broad-line region. Even though there are archival data of our source from the XMM-Newton mission, the monitoring frequency is insufficient to perform a study of short-term X-ray variations. Therefore we cannot infer whether similar behavior occurs in Mrk 477. Nevertheless, we cannot rule out this possibility. 

Another possible explanation lies within the structure in the dusty torus, which could provide observational windows to observe the inner parts of the otherwise obscured AGNs. The original unified scenario introduced by \citet{ant} includes a geometrically thick obscuring rotating region, or in other words, a torus. It is worth noting that \citet{kro} suggested that such a structure would be hard to maintain and introduced a clumpy structure into the model of the torus. The model introduced by \citet{nen} counts no more than five or ten clouds, which are unsuitable to explain high-frequency variations. Considering orbital motion for the individual clouds in the dusty torus, variability is expected whenever these clouds, temporarily, do not completely hide the central engine. This may be the case for a few Seyfert galaxies that have shown spectral variations, changing from type 2 to type 1 AGNs and vice versa. However, these variations occurred on timescales of years, and our data do not allow us to observe such a behavior.

\section{Conclusions}

Microvariation in type 2 quasars is a phenomenon previously unaccounted for. From our observing campaign, we reported a detection of one variable (Mrk 477) and one unconfirmed variable type 2 quasar (J0759+5050) that were detected as variable during one observing block shorter than two hours. Both results were obtained using a statistical method, namely a one-way analysis of variance, ANOVA, which has previously proved to be a powerful test for the detection of microvariations in AGN light curves. Given the observation history of the variable object Mrk 477, we discussed the  possible causes of the variations. Our observing campaign was limited to one filter in the optical regime, therefore we cannot confirm any of our hypotheses by observational evidence. Because of the possible importance of microvariability to the study of the physics of the type 2 quasars and their obscured regions, it is worth investigating whether this is a ubiquitous phenomenon in this class of AGNs.

\begin{acknowledgements}
The observations reported here were acquired at the Observatorio
Astronómico Nacional in the Sierra San Pedro Mártir (OAN-SPM), B.C.,
Mexico.
This research has been supported by the Spanish Ministerio de Econom\'ia y Competitividad (MINECO) under the grant AYA2014-58861-C3-1.
IO acknowledges support from the European Research Council (ERC) in the form of Advanced Grant, {\sc cosmicism}.
JAD is grateful for the support from the grant UNAM-DGAPA-PAPIIT IN110013 Program and the Canary islands CIE: Tricontinental Atlantic Campus. 
AE acknowledges support by the grant AYA2012-30789.
J.I. Gonz\'alez-Serrano is grateful for the support from  AYA2011-29517-C03-02.
The authors also would like to thank our colleague Yair Krongold for useful suggestions.
\end{acknowledgements}

\clearpage

\end{document}